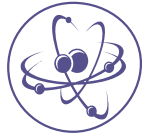

## Journal of Computer Engineering & Information Technology

a SciTechnol journal

**Research Article**

# A New Approach for Image Authentication Framework for Media Forensics Purpose

**Ahmad M Nagm[1]\*, Khaled Y Youssef[2] and Mohammad I Youssef[1]**

## Abstract

With the increasing widely spread digital media become using in most fields such as medical care, Oceanography, Exploration processing, security purpose, military fields and astronomy, evidence in criminals and more vital fields and then digital Images become have different appreciation values according to what is important of carried information by digital images?. Due to the easy manipulation property of digital images (by proper computer software) makes us doubtful when are juries using digital images as forensic evidence in courts, especially, if the digital images are main evidence to demonstrate the relationship between suspects and the criminals. Obviously, here demonstrate importance of data Originality Protection methods to detect unauthorized process like modification or duplication and then enhancement protection of evidence to guarantee rights of incriminatory. In this paper, we shall introduce a novel digital forensic security framework for digital image authentication and originality identification techniques and related methodologies, algorithms and protocols that are applied on camera captured images. The approach depends on implanting secret code into RGB images that should indicate any unauthorized modification on the image under investigation. The secret code generation depends mainly on two main parameter types, namely the image characteristics and capturing device identifier. In this paper, the architecture framework will be analyzed, explained and discussed together with the associated protocols, algorithms and methodologies. Also, the secret code deduction and insertion techniques will be analyzed and discussed, in addition to the image benchmarking and quality testing techniques.

## Introduction

In the last decade, a wide spread of digital imaging is witnessed as a normal consequence of the digital revolution era. The imaging capabilities become associated to a low cost mobile device that makes the image capture of any event is a simple, easy and cheap task in comparison to the last decade. On the security aspect, the manipulation and the change of the digital image become available using a proper photo edit application that could be embedded on the image capturing as well. The security aspect is not only restricted to post capture edit by an attacker but it also includes the real time attack of the image during capture on a security compromised IP camera system.

---

*Corresponding author: Ahmad M Nagm, Electrical Engineering, Faculty of Engineering, Al Azhar University, Egypt; Tel: +20 102 722 7142; E-mail: ahmadnagm@gmail.com

However, the approach discussed in this paper is based creating an identity to the capture device and link it to the captured image that is distorted on any trial to change the image. Accordingly, the framework authenticates the image against any potential attacks for modification [1]. In this paper, a framework is presented including solution architecture, suite of algorithms and protocols that govern the authentication process and verification of the captured image. The work is supported by complete simulation model. As a result, Image authentication framework detect originality effectively against attackers and malicious penetrations and consequently it prevents destroying or modification of information content that is carried by digital image [2].

In addition, the paper shall discuss the methodology to authenticate image originality by hiding print signs "secret code" in a color images using steganography approaches together with appropriate encryption symmetric algorithms for hide secret messages as a protection patterns [3]. The output RGB images of the processing must be originality protected RGB images where the Proposal Approach is based on combination of Crypto-Steganography-Hashing and effectively enhances the strength power of Photo as forensic evidence. Where the approach present novel method by creation a unique secret code as a 'fingerprint' or 'digital signature' embedded into RGB images [4].

Security algorithms used in the proposed technique are the Advanced Encryption Standard (AES) symmetric algorithm and Secure Hash Function (SHA) with two mixed of steganography methods at space domain and spread spectrum frequency domain executed on one component or mixed of more components from input RGB images [5] to Originality protection for carried data of images.

### Related work

Most of the research work done on image authentication usually study authentication with respect to the originality of the image content, that is the modification of the content such us exchanging the face of the person with someone else or removing an image object, should be detectable. Those approaches are associated with a probability of success that could be very high but does not reach one. So we mention methodology of execution this approach based on used articles at another topics.

- Integrity Protection

- Cryptography

- Steganography

**Integrity protection:** From ancient most used methods of integrity protection to save information from altering and modification by malicious adversary [6] are cryptographic hash functions. A cryptographic hash function: is a sub-section of hash function has a certain properties distinguished it to the best for using in cryptography, this mathematical algorithm map data(message) as a input with arbitrary size to a bit string have fixed size as an output(Digest) formed as a hash function, so no way to get the input data from an output hashed value(Digest) else attempt a brute-force operation to find all possible inputs which any it's result is matched because a function is infeasible to invert this method is referred to one-way property [7] as depicted in equation below.

$$h = H (M) \qquad (1)$$

where: H is hash function, M is input variable-length block of data and h is fixed-size hash value.

**Cryptography:** AES is a symmetric block cipher and there are some versions of AES have different keys of 128,192, 256 and 512 bits, AES have four basic function are: add key, Nibble substation, Shift raw and Mix column as shown in Figure 1 below, operation in AES algorithm are XOR, S-Box (8×8), and shift left rotate, and multiplication. The arithmetic Operations of addition, multiplication, and division are performed over the finite field (GF ($2^8$)) defined with this polynomial.

$$M(x) = x^8 + x^4 + x^3 + x + 1 \qquad (2)$$

Where: M(x) is irreducible polynomial.

**Steganography:** Cryptography protect the content of messages on the other hand, cryptography scrambles a message in such a way that it cannot be understood while steganography concerned to conceal their very existence of messages [8] the presence of hidden information is revealed or suspected.to avoid some weaknesses of encryption method must use steganography techniques where appearance of cipher-text in flow traffic lead attacker to access original content by breaking it, if he cannot decrypt data may destroy it. Although data unreadable but still exists as data. If given enough time, someone could eventually decrypt the data. So we try to concealment existence encrypted data. Steganography and cryptography are integrated to robust security model to conceal existence and encrypt information from malicious intruder and attacks but neither technology alone is perfect. For more optimal security robust then combine steganography methods with encryption algorithms by specific protocols, where both the data hiding and encryption techniques [9] are found to be the main mechanisms in data security. Steganography equation is 'Stego-medium = Cover medium + Secret message + Stego-key      (3) where: Stego-medium: is a final form including cover medium which embedded secret message and stego-key. Cover-medium: is the medium which conceal existence of secret message like a cover-text, cover-Image, Cover-Audio, cover-video. Secret message: is the hidden message. Stego-key: is used to hide message inside cover-medium at sender and recover the test at receiver side.

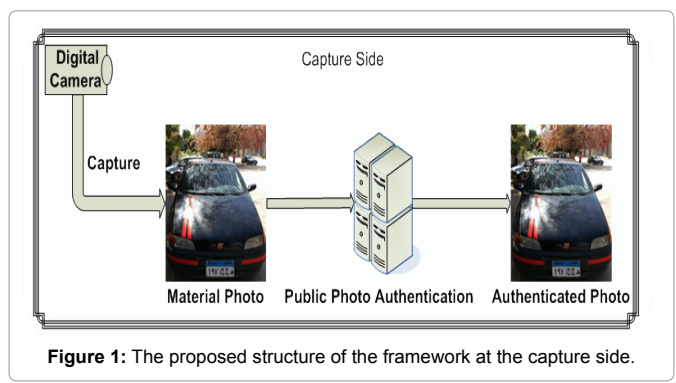

**Figure 1:** The proposed structure of the framework at the capture side.

## Image steganography

Encoding Secret Messages in cover-Images is the most using technique because it depends on poor sensitivity of high frequencies for human visual system (HVS). In Suresh Kumar [10], research suggested new model algorithm which "aimed to embed text file into color image by replacing value of one component (R, G and B) of image pixel by one bit of text form left to right again the same previous process for next seven bit (LSB) method of steganography. In Chandreyee Maiti [11] research present a new model which aimed to hiding secret message based on LSB substitution method by using one of three difference steganography methods the first is by using the least significant bit Second, by using the last two significant bits and the last is by using diagonal pixels of the image. Before steganography method performed we used asymmetric stream cipher algorithm (RSA) to encrypt secret message [12], where secret message was text and cover medium was color or gray image. In M.RAJKAMA [13] research paper a secured robust approach of embedded secret message in cover-medium to obtained stego image and then encrypted stego-image was proposed. Where the secret message was encrypted by asymmetric block cipher algorithm (RSA) and inserted in cover-medium(image)based on hash-LSB technique to get stego-image after that the stego image was encrypted by Blowfish algorithm and then sent this ciphered data (encrypted stego-image). In Manjula [14] research paper presented novel method in embedded secret message into cover-image (RGB image) based hash 2-3-3 LSB substitution approach. Where exacted one byte=8 bits of secret message at a time and insert them into cover RGB image as: Divided the secret message into the most 2 significant bits, the following 3 significant bits and the least 3 significant bits-Exacted 4LSB of each RGB components of cover-image, using hash function to find the location for inserting the secret data into cover image, embedded the secret message (8bits exacted) into cover-medium (RGB image) according to previous specified location in order to 2, 3, 3 respectively.

## Proposed model

Proposed System Architecture: In the send phase, we study the performance of seeding print signs into RGB images with Quality of the image. On the other word, make the image have Originality Protection by seeding print signs without effect on image Quality as possible via framework at Figure 2 below by using mathematical and human visual system metrics.

In the send phase, we received digital forensic via internet, so we want this digital forensic infected by active attack or not, where we test performance detection of proposal architecture, where the client ask Web server Front end to establish session with PAS server to execute the test on received media forensic by Run One of proposal Algorithm as we mention later. The architecture is based on two modes of operation; the system can work with namely:

A. Offline mode of operation: In this mode, the camera is not a must to be IP or network camera, but it is essential to have a figure print of each camera that will be used through public photo authentication (PPA) protocol to check and verify that the photo is original and not manipulated using photo manipulate application. The camera should be equipped with a unique identity (serial could be identity the camera). In the offline operation mode, the user just verifies an image under investigation regardless of the source and without any information on the author. The user request the verification service directly from a front-end web service that enables back to back verification with the PAS server that should be located in a different DMZ zone for security purpose.

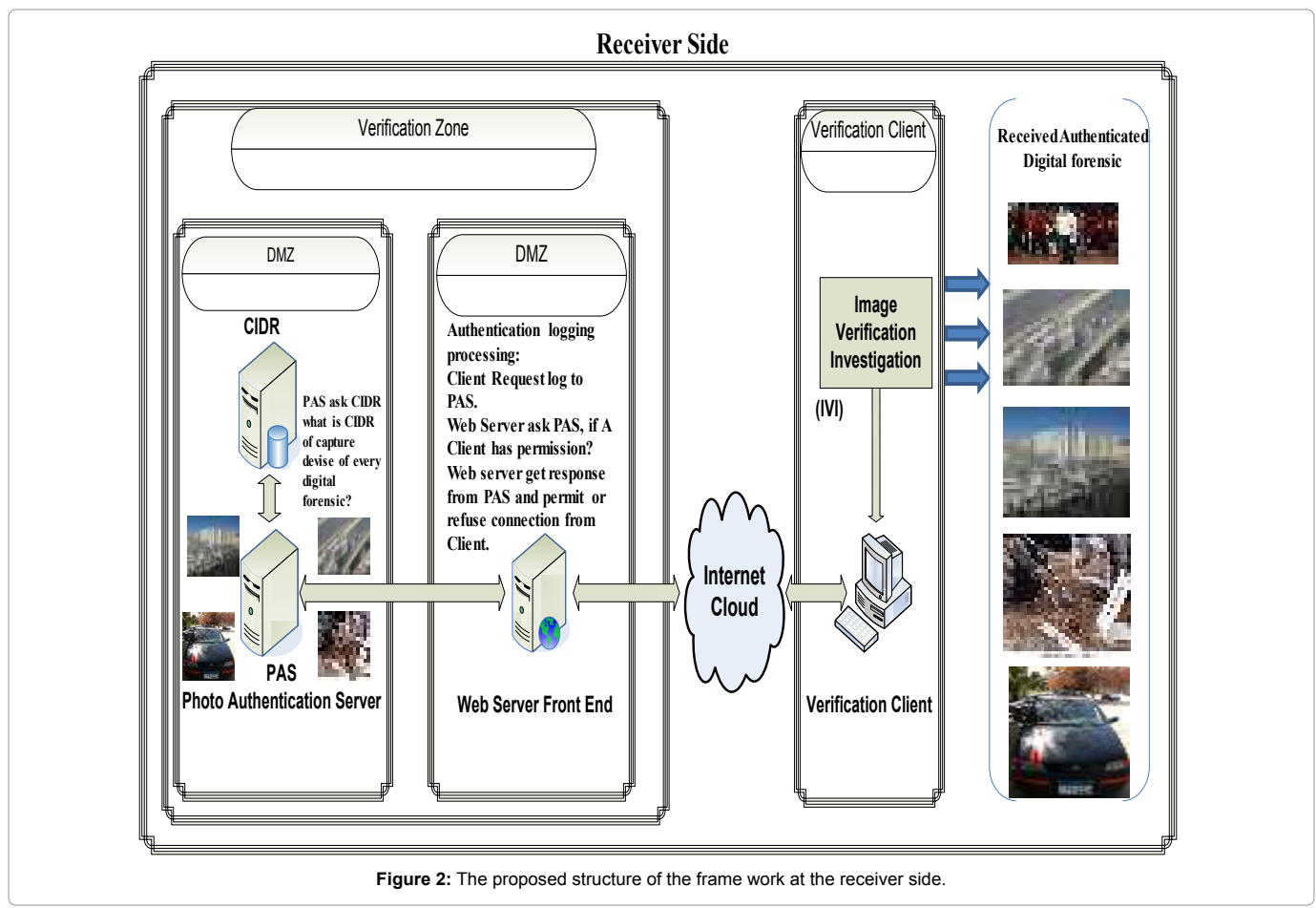

**Figure 2:** The proposed structure of the frame work at the receiver side.

B. Online mode of operation: In this mode, the camera is an IP or network camera that should be equipped with camera MAC address and ID. The online mode of operation has three authentication modes:

- **Public Authentication Mode:** In this mode, the camera ID exists is an identifier on any registered image capture device that exist on both the equipment side (hardcoded on camera) as well as the authentication server register side known as Camera ID Register (CIDR) . The image capture device ID is thereafter transformed into photo ID using a convenient hashing algorithm which is the secret code word proposed to be embedded and hidden in the image post capture and is recognized as a forensic evidence for originality of image. In this mode of authentication, the verification party will send the image under investigation to the Photo Authentication Server (PAS) that will in turn do the following:

- Extract the photo ID from the image

- Query the CIDR on the photo ID to get the corresponding camera ID

- Apply the technique standard hashing algorithm on the image using the camera ID and deduce the photo ID. The above steps are governed and detailed in the Public Photo Authentication Protocol (PPAP) proposed.

- **Private authentication mode:** In the private authentication mode, the proposed framework reserves a space for private image communications for enterprises and closed groups. In this mode of authentication, the users can verify the image using a secret code that result from applying a hashing algorithm on the image. In this case, the algorithm will not use neither camera ID not the PPAP protocol but just an initial verification of the image within an organization using series of keys known for the organization. The disadvantage of this mode, that it could not be considered as an official model legally and thus cannot be used as an evidence proof for forensic techniques.

- **Hybrid authentication mode:** In this mode, both authentication modes are used according to the importance of the image. Critical images should be stamped with a secret code and authenticated using photo authentication modes while normal and temporary images could be authenticated using private authentication mode. In this mode of authentication, the images could be classified according to standard classification techniques into permanent images and temporary images. The former should include all images that will be archived for future use and could be evidence in the future while the latter is the images family that is captured for temporary use and usually deleted after a time.

### Photo authentication methodology

In this section, the algorithm of deducing, and implanting a secret code in a captured image will be discussed. The extracted code is characterized by the following features:

- The code is a function of the image details, colors, timestamp, and image acquisition device ID.

- The code cannot be reverse decoded as it is a result of applying a one-way hashing function [6].

- The code has a minimal impact on capturing latency performance as well as image quality. The benchmarking technique will be discussed on the next section.

- The secret code validation time is controlled by being done though light algorithm, and efficient protocol.

The photo authentication algorithm is the security algorithm concerned with integrity protection of the image in a sense that indicates also the originality of the image (i.e. the image being investigated is the image created by the first author image acquisition device).

The algorithm framework is divided into two main algorithms: As explained in Figure 3, in this phase it is designed to protect

RGB images originality by embedding secret originality identifier (SOI) into color captured images distributed over the whole image according to algorithm rules.

Algorithm Steps are explained as follows:

- **Post-capture stage:** in this stage, the captured image is prepared for processing, in a memory block, and extract image main characteristics (e.g. dimensions). In addition, the image is decomposed into its main basic color components (Red, Green and Blue). The components are then resized for further processing stage.

- **Modifier selection stage:** in this stage one color component is selected as a modifier another one component is selected as modified while the third one is left unprocessed.

- Integrity protection Stage: execute Cryptographic Hash Function by Secure Hash Algorithm at manufacture information of Camera's SHA (Camera ID) to obtain K (Photo ID) that will be amended to the image [7].

- **Encryption stage:** Advanced Encryption Standard AES

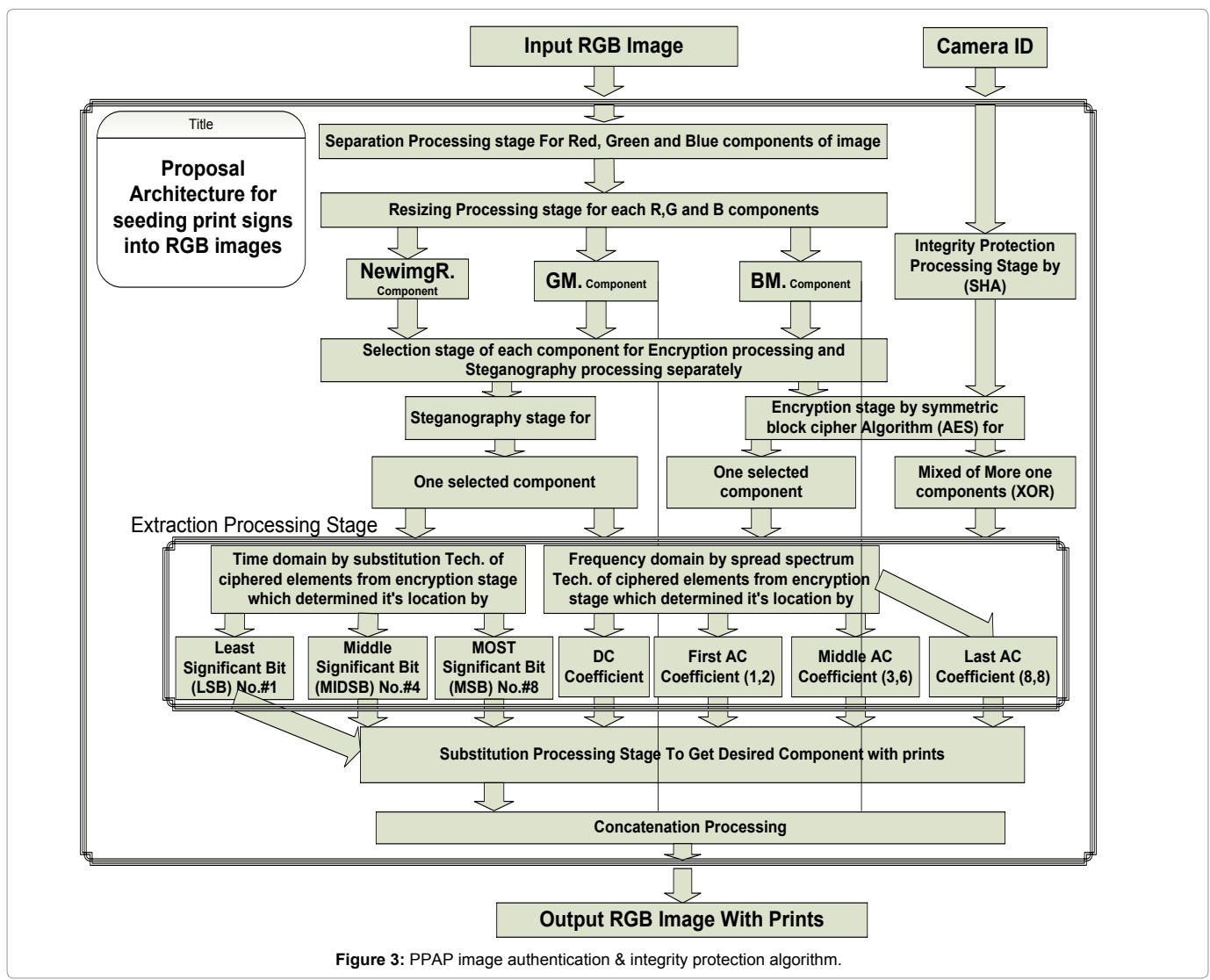

**Figure 3:** PPAP image authentication & integrity protection algorithm.

is used with the key K1 (camera ID) [15] on the modifier color component to produce encrypted modifier color component [16].

- **Frequency transform:** in this stage, a DCT is used to transform the modifier and modified color component into frequency domain for further processing.

- **Bit substitution stage:** this step we execute it at space domain [8] by substitute "LSB or MSB or Bit no. #4 for every byte" [9]. Or performed at frequency domain, the modifier and modified matrix components are organized into 8x8 blocks, then middle AC coefficients (3,6) or DC (1,1) or First AC (1,2) or last AC coefficients (8,8) is substituted as well as between modifier and modified blocks [10,17].

- **Frequency inverse transform:** in this stage, the results of substitution are inversed back again to space domain.

- **Integrity code implantation:** in this stage, the final modified color component replaces the original modified color component in the original color image.

As explained in Figure 4, in this phase it is designed to detect color images originality by checking secret originality identifier (SOI) existing color captured images.

Algorithm Steps are explained as follows:

- **Post-receive Stage:** in this stage, the modified and modifier color component is extracted from the received image.

- **Modifier processing stage:** in this stage the modifier component are encrypted using Advanced Encryption Standard Algorithm (AES-128) with the key k1 that is retrieved using PPAP protocol given the photo ID K extracted from the image.

- **Modifier frequency transform:** in this stage, DCT is applied on the final modifier component and the modified color component in 8x8 blocks format.

- **Integrity verification stage:** in this stage, in case of space domain [11] the selected "LSB or MSB or Bit no. #4 form every byte" of standard image [12]. Or in case of frequency domain, the selected coefficients namely "(1,1), (1,2), (3,6) and (8,8) for every 8*8 block" are compared against each other correspondingly in the modifier image color component against the modified image color component. Accordingly, the integrity is proven and originality is stamped for the image under investigation [13,14].

## Results and Discussions

The simulation strategy is based on two main approaches, the first approach to assess the impact of the change on the space domain and frequency domain while the second approach is based on simulating active attacks and assess the immunity of the system against several attacks attempts. The simulation tool used is MATLAB Version 8.5.0.197613 (R2015a) applied on many of digital forensic as a color images. RGB images A, B and C are represented by 24 bit, captured in JPEG format, where the Image A has size "152*400*3", the Image B has size "312*522*3" and the Image C has size "800*600*3" as shown in Table 1 used for analysis and testing the experimental scenarios that apply the aforementioned algorithms.

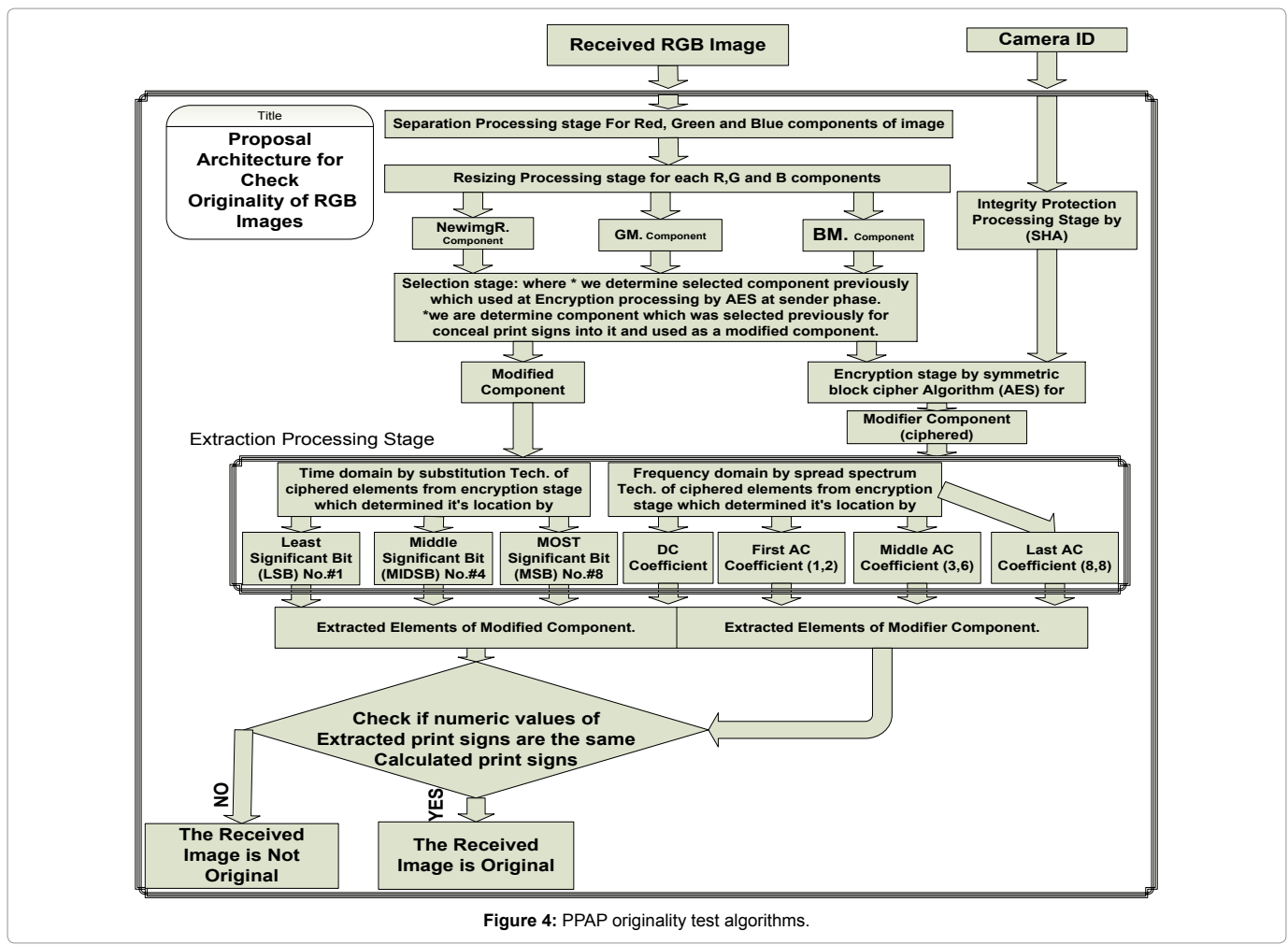

**Figure 4:** PPAP originality test algorithms.

A. In this section, we will compare the images before applied the algorithm and after the algorithm and accordingly the visual assessment of the impact for the normal user. In the coming section, a quantitative analysis is held to assess the impact of the change.

B. Space Domain Scenarios

**Group 1: Originality steganography using single color modifier**

- **Scenario 1:** In this scenario, the forth bit in the ciphered modifier pixel component (blue color component) substitute the corresponding forth one in the unchanged modified pixel component (red color component) as mention previously.

- **Scenario 2:** In this scenario, the least significant bit in the ciphered modifier pixel component (blue color component) substitutes the corresponding least significant one in the unchanged modified pixel component (red color component).

- **Scenario 3:** In this scenario, the most significant bit in the ciphered modifier pixel component (blue color component) substitute the corresponding most significant one in the unchanged modified pixel component (red color component) as mention previously.

**Group 2: Originality steganography using dual color modifier**

- **Scenario 4:** Originality protection performed by substitute the XOR result of dual basic color components as a modifier component to modified component for every byte at color image.

**A. Frequency domain scenarios**

**Group 1: Originality steganography using single color modifier:** In the frequency domain manipulation, captured image is transformed using DCT, ciphered and the resultant matrix is proposed accordingly in the coming 5 scenarios

- **Scenario 1:** In this scenario, substitution process is performed at frequency domain after transform ciphered modifier and modified component, by Last AC Coefficient (8, 8) of every 8*8 block at color images.

- **Scenario 2:** In this scenario, substitution process is performed at frequency domain after transform ciphered modifier and modified component, by Last DC Coefficient (1, 1) of every 8*8 block at color images.

- **Scenario 3:** In this scenario, substitution process is performed at frequency domain after transform ciphered modifier and modified component, by First AC Coefficient (1, 2) of every 8*8 block.

- **Scenario 4:** In this scenario, substitution process is performed at frequency domain after transform ciphered modifier and modified component, by middle AC Coefficient (3, 6) of every 8*8 block at color images.

**Group 2:** Originality protection performed by using the result

| Image A (Before seeding secret code) | Image B (Before seeding secret code) | Image C (Before seeding secret code) |
|---|---|---|
| 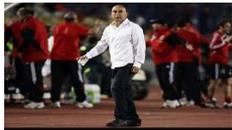 | 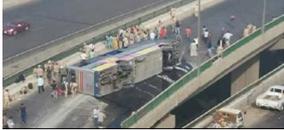 | 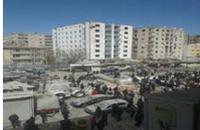 |
| Image A (After seed secret code at Least Significant Bit for every byte of input image) | Image B (After seed secret code at Least Significant Bit for every byte of input image) | Image C (After seed secret code at Least Significant Bit for every byte of input image) |
| 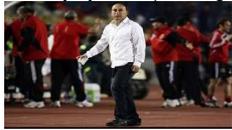 | 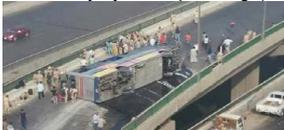 | 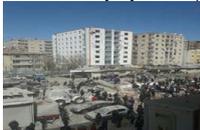 |
| Image A (After seed secret code from XOR of dual component at LSB of input image) | Image B (After seed secret code from XOR of dual component at LSB of input image) | Image C (After seed secret code from XOR of dual component at LSB of input image) |
| 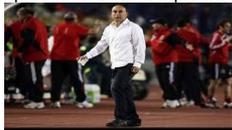 | 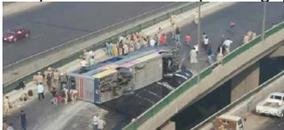 | 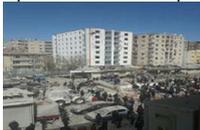 |
| Image A (After seed secret code at Middle AC Coefficient from XOR of dual component at evey 8*8 block of input imag) | Image B (After seed secret code at Middle AC Coefficient from XOR of dual component at evey 8*8 block of input image) | Image C (After seed secret code at Middle AC Coefficient from XOR of dual component at evey 8*8 block of input image) |
| 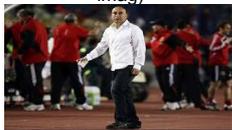 | 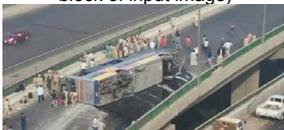 | 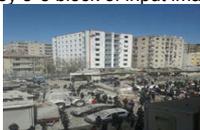 |

**Table 1:** It shows performance of applying proposal scanners on quality of color digital images.

of dual basic color components at frequency domain as a modifier component.

- **Scenario 5:** In this scenario, substitute the Middle AC Coefficient (3,6) at frequency domain of XOR result for dual Components for 8*8 block as a modifier component into transformed modified component at frequency domain then perform Invers transform processing to get protection originality color images at space domain.

With "Camera ID = acef "the output was (Photo ID) K= 86E152C142DB1256FC1EF004ADEB7B935741D94D, in the coming section, a quantitative analysis is held to assess the impact of the change.

### Assessment of simulation results

We estimate performance of applied Scenarios on RGB Image Quality duo to embedding Secret Originality Identifier at Image Authentication & Integrity protection phase by using the objective IQM assessment methods depending on Mean Absolute Error, Mean Square Error and Peak Signal to Noise Ratio as a mathematical metrics and Structure Similarity Index and Universal Image Quality Index according to Human Visual System metrics, and then at Originality Test phase we will test performance of Manipulation Detection.

### Impact assessment using error mathematical metrics

When we assessment the output of digital forensic duo to applied proposal scenarios according to image quality measurements, as:

**Mean absolute error,** defined by:

$$MAE = 1/N \sum_{(i=1)}^{N} |x_i - v_i| \qquad (4)$$

**Mean Square Error (MSE),** according to:

$$MSE = 1/(M*N)* \sum_{(i=1)}^{m} \sum_{(j=1)}^{n} [x(i,j) - v(i,j)]^2 \qquad (5)$$

**Peak signal to Noise Ratio (PSNR)** [18], defined by:

$$PSNR = 10 \; [log]\_10 ((R^2 / MSE)) \qquad (6)$$

Where N, M are the dimension of matrix and x, v are variable value at point i, j and R is maximum signal value that exists in our original image, R = 255 for an 8-bit image.

**Impact assessment using human visual system metrics**

The Human Visual Based on Measurement [19-22] based on metrics:

**Universal Image Quality Index (UIQI):** we measure image quality distortion [23]:

Loss of correlation, luminance distortion, and contrast distortion.

$$Q = (\sigma_{xv} / \sigma_x \sigma_v)*((2*x^- *v^-)/(x^-{}^2 + v^-{}^2))*((2*\sigma_x*\sigma_v)/(\sigma_x{}^2 + \sigma_v{}^2)) \qquad (7)$$

**Structural Similarity Index (SSIM):** we measure the similarity between original image and image with print signs, and then we calculate two means and two standard derivations and one covariance value are computed for the images as:

$$SSIM(X,V) = (2^{*-}x{}^{*-}v + c1)*((2*\sigma_{xv} + c2)) / \\ ((^-x{}^2 + {}^-v{}^2 + c1)*(\sigma_x{}^2 + \sigma_v{}^2 + c2)) \qquad (8)$$

Where: $\bar{x}$, $\bar{v}$ are mean for $x$ and $v$ c1 and c2 are variables to stabilize the division with weak denominator; c3=c2/2. From Table 1 and Figure 5, the convergence of input RGB images (A, B and C) duo to applied proposal scenarios and substitution processing are executed at space domain according to: Impact on MAE: was the scenario 2 and 4 have the best results and approximate effect not be mentioned, for scenario 1 the effect was more but unnoticeable by human eye but with applying scenario 2 the result is very bad and the output color images were more noisy. From Figure 6 was the least effect on image quality for scenario 2 and scenario 4 and no significant effect for scenario1, but scenario3 achieve the most effect at output color images quality According to Impact on PSNR, from Figure 7 duo to applied scenario2 and scenario4 have the beast convergence of output images then scenario 1 have acceptable results, but scenario3 was the bad results. From Table 1 and Figure 8, the convergence of input RGB images (A, B and C) duo to applied proposal scenarios

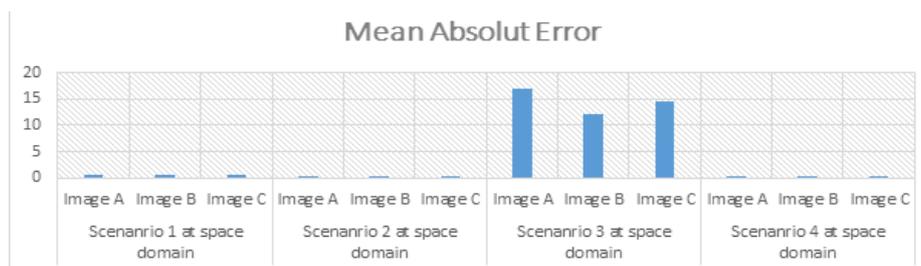

**Figure 5:** Shows convergence using the mean absolute error of output images duo to applied proposal scenarios at space domain.

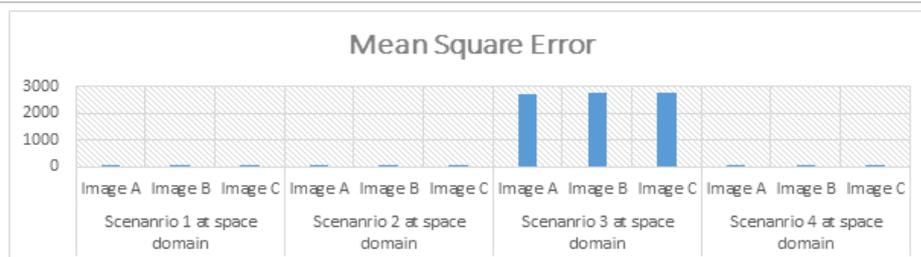

**Figure 6:** Depict convergence using the mean square error for output images duo to applied proposal scenarios at space domain.

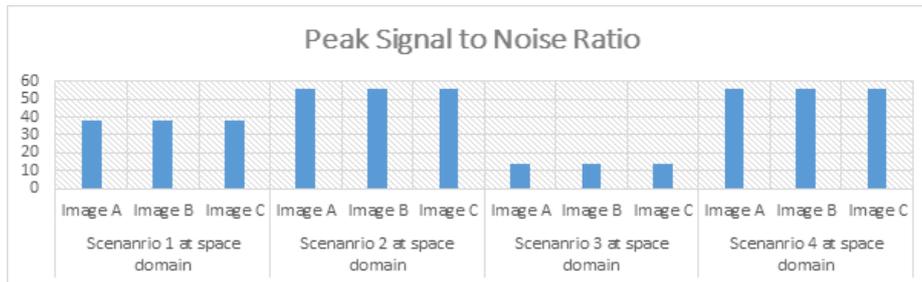

**Figure 7:** Shows convergence using the peak signal to noise ratio for output images duo to applied proposal scenarios at space domain.

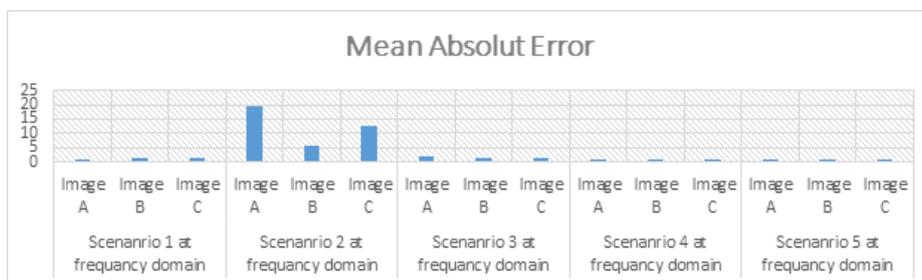

**Figure 8:** Shows convergence using the mean absolute error of output images duo to applied proposal scenarios at frequency domain.

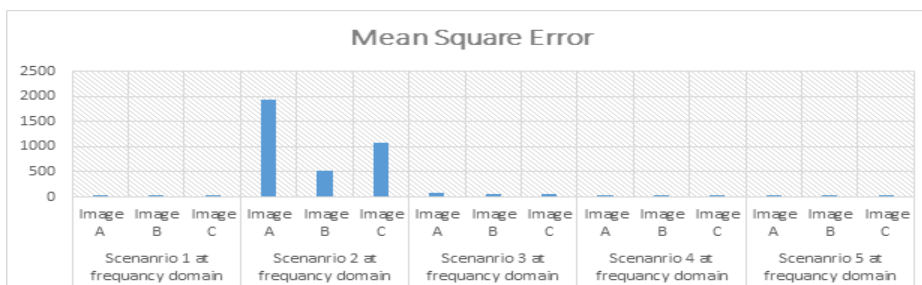

**Figure 9:** Depict convergence using the mean square error for output images duo to applied proposal scenarios at frequency domain.

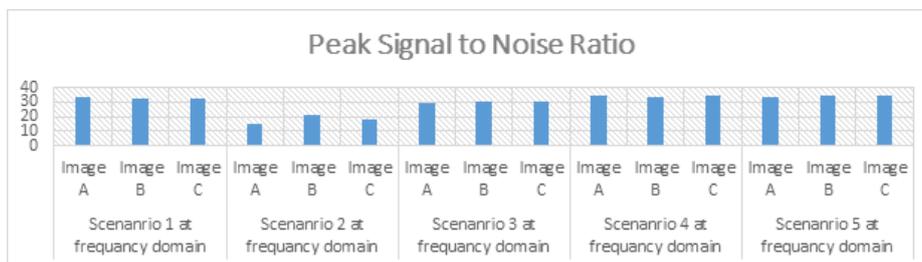

**Figure 10:** Shows convergence using the peak signal to noise ratio for output images duo to applied proposal scenarios at frequency domain.

and substitution processing are executed at frequency domain according to: Impact on MAE was the scenario 4 and 5 have the best results and approximate effect not be mentioned, for scenario 1 and scenario 3 have the effect was more but unnoticeable by human eye but with applying scenario 2 the result is very bad and the output color images were very noisy. From Figure 9, the image quality depend on Impact of Mean Square Error duo to applied proposal algorithms and execute substitution on frequency domain was the least effect on image quality for scenario 4, scenario 5 and scenario

1 respectively and no significant effect for scenario 3, but scenario2 achieve the most effect at output color images quality and the output were very noisy. According to Impact on PSNR, from Figure 10 the least effect for output image quality and more convergence were for scenario 4, scenario 5, scenario1 and scenario3 respectively, but scenario2 were the bad results and detect the noise by human eye. From Table 1 and Figure 11, the correlation of input objects RGB images (A, B and C) duo to applied proposal scenarios and substitution processing are executed at space domain according to: Structure

Similarity Index, was the scenario 2 and 4 have the best results and approximate effect not be mentioned, for scenario 1 the effect was more but unnoticeable by human eye but with applying scenario 2 the result is very bad and the output color images were more noisy. From Figure 12, the correlation of input objects RGB images (A, B and C) duo to applied proposal scenarios and substitution processing are executed at space domain according to Universal Image Quality Index, was the scenario 2 and 4 have the best results and approximate effect not be mentioned, for scenario 1 achieve good results and effect unnoticeable by human eye but with applying scenario 2 the result is very bad and the output color images were more noisy. From Table 1 and Figure 13, the correlation of input objects RGB images (A, B and C) duo to applied proposal scenarios and substitution processing are executed at frequency domain according to Structure Similarity Index, was the scenario 2 and 4 have the best results and approximate effect not be mentioned, for scenario 1 the effect was more but unnoticeable by human eye but with applying scenario 2 the result is very bad and the output color images were more noisy. From Figure 14, the correlation of input RGB images (A, B and C) duo to applied proposal scenarios and substitution processing are executed at frequency domain according to Universal Image Quality Index was the scenario 4, scenario 5, scenario 1 and scenario 3 have the best results respectively and approximate effectiveness on output image quality not be mentioned, but with applying scenario 2 the result is very bad and the output color images were more noisy by human eye.

## Assessment of Framework Vulnerability Against active attacks

The vulnerability assessment approach is based on assuming an attack on one of the quadrants zones shown in figure below. In this approach, the technique should show full immunity against active attacks, whereas the image could be changed in any of the areas explained. Moreover, the assessment technique shall include as well the color component changes since the algorithm is based on manipulation based on color components. let take RGB image f has dimensions 344*512 pixel was represented by 8 bits for every main color component with JPEG format, where we conceal some evident of criminal and then apply the approach algorithms to emphasis the Modification Detection occur as shown in below (Figure 15).

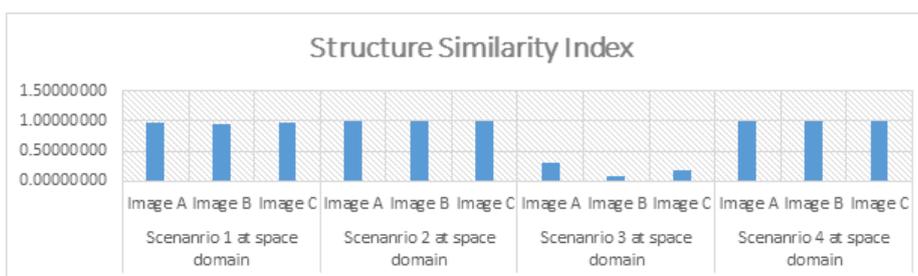

**Figure 11:** Shows correlation using structure similarity index metrics for output images duo to applied proposal scenarios at space domain.

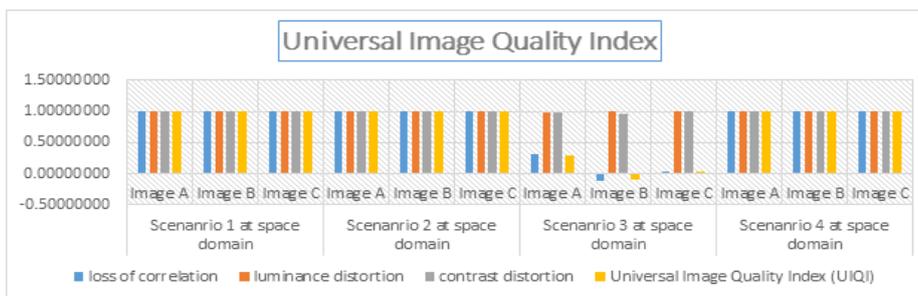

**Figure 12:** Depicts image quality using universal image quality index metrics of output images duo to applied proposal scenarios at space domain.

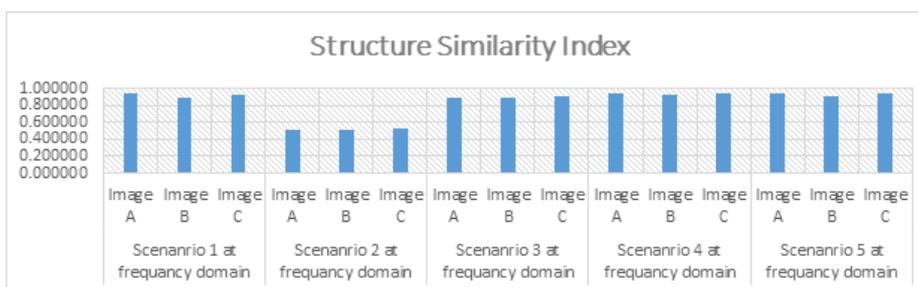

**Figure 13:** Shows correlation using structure similarity index metric for output images duo to applied proposal scenarios at frequency domain.

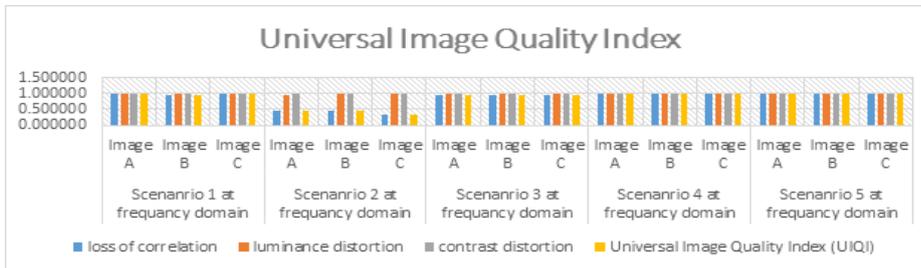

**Figure 14:** Depicts image quality using universal image quality index metrics of output images duo to applied proposal scenarios at frequency domain.

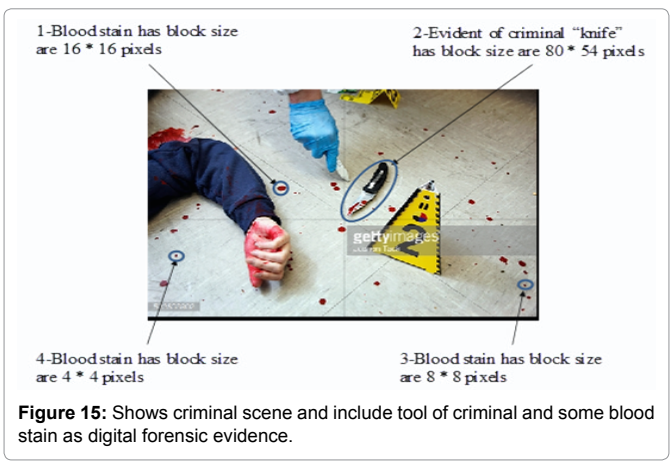

**Figure 15:** Shows criminal scene and include tool of criminal and some blood stain as digital forensic evidence.

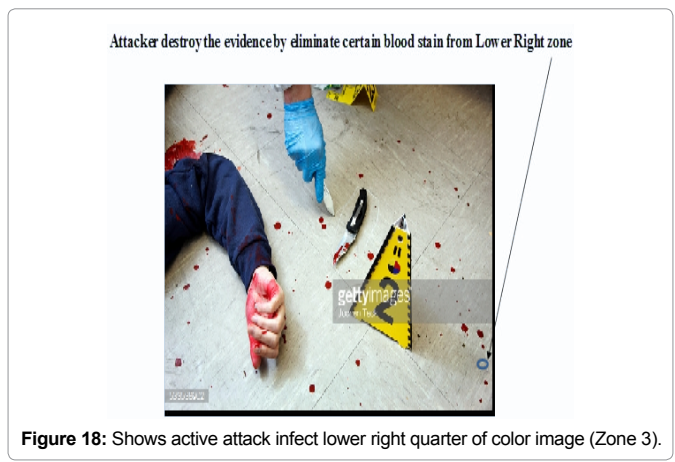

**Figure 18:** Shows active attack infect lower right quarter of color image (Zone 3).

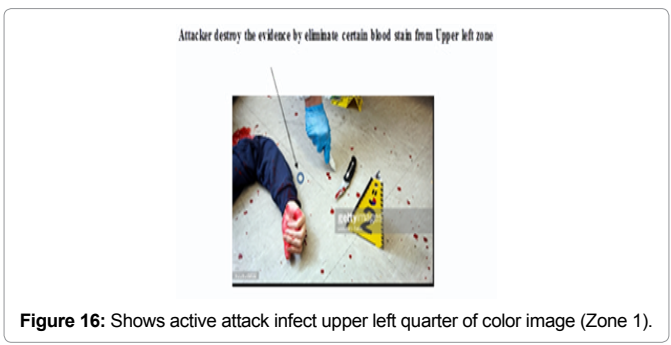

**Figure 16:** Shows active attack infect upper left quarter of color image (Zone 1).

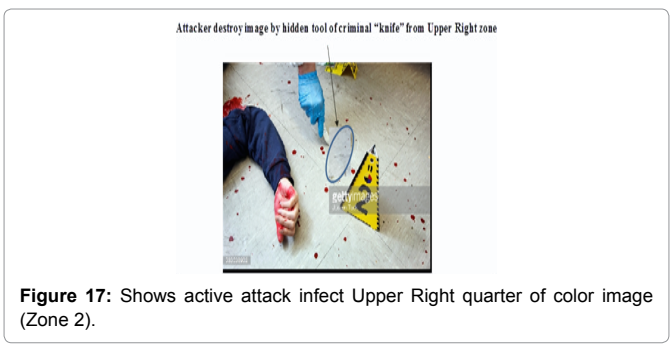

**Figure 17:** Shows active attack infect Upper Right quarter of color image (Zone 2).

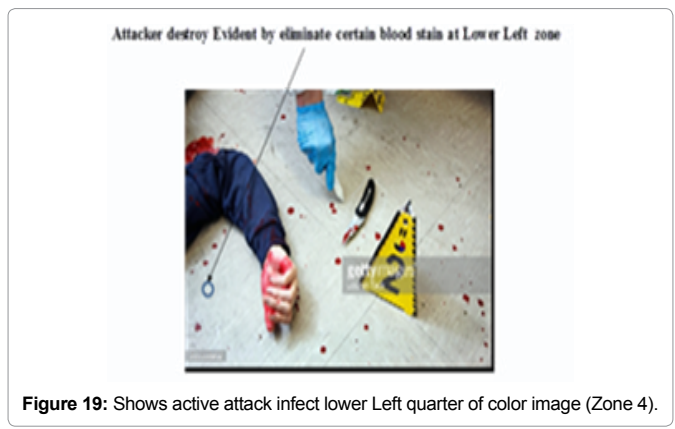

**Figure 19:** Shows active attack infect lower Left quarter of color image (Zone 4).

From previous experiments we note the approaches algorithm detect active attack with differences sizes from 80*54 pixels, 16*16 pixels, 8*8 pixels and 4*4 pixels even if manipulated size are less size because the algorithm measure correlation between color images depending on embedded secret code at every pixel which changing with any modification for pixel at any location of color images.

## Conclusion

From previous results we can deduce the Scenario2 and Scenario 4 have the best results and no effectiveness on image quality duo to substitute secret code at LSB "bit number 1 for every byte" from modifier (ciphered Blue) component at modified (Red) component even if, production modifier component was by single or result of XOR dual component then substitute by "bit number 4" and the least results achieve when substitute by MSB "bit number 8", In case of substitution at frequency domain the Scenario4 and Scenario5 have the best mean that no significant on output image quality duo to substitute by Middle Ac coefficient "element number (3, 6) for every 8*8 block" even if, the production of modifier component were by single or result of XOR dual components then substitute by last AC coefficient "element number (8,8)" has less effect on image quality then first AC coefficient "element number (1, 2)" and the DC coefficient

"element number (1,1)" was the bad effect and output images more noisy to detected by human eye. The steganography based approach to hide the evidence of originality of the photos on creation is vital nowadays digital Era that is characterized by difficulties to distinguish photos based on integrity or credibility. The problem severity appears more in photos used in official and legal conditions. One solution is to ensure no physical access to the camera, but the challenge is still there if the camera is an IP camera. The simulation results proved that the image could be integrity protected with minimal unobserved impact on the quality of image if the image processing stage posts the capturing stage include the huddling of integrity code including the ID of the camera in the image that becomes the image that we can consider it digitally the original copy (Figures 16-19).

## Author Affiliation                                    Top

¹Electrical Engineering, Faculty of Engineering, Al Azhar University, Egypt

²Communications Engineering, Faculty of Space Technology, BSU University, Egypt